\documentclass[acmsmall,screen,nonacm]{acmart}
\title{GeckoGraph: A Visual Language for Polymorphic Types}

\AtBeginDocument{%
  }

\setcopyright{none}
\renewcommand\footnotetextcopyrightpermission[1]{} % removes footnote with conference information in first column
\author{Shuai Fu}
\email{shuai.fu@monash.edu}
\affiliation{%
 \institution{Monash University}
 \country{Australia}
}

\author{Tim Dwyer}
\email{tim.dwyer@monash.edu}
\affiliation{%
 \institution{Monash University}
 \country{Australia}
}

\author{Peter J. Stuckey}
\email{peter.stuckey@monash.edu}
\affiliation{%
 \institution{Monash University}
 \country{Australia}
}

\begin{document}

\begin{abstract}
 Polymorphic types are an important feature in most strongly typed programming languages. They allow functions to be written in a way that can be used with different data types, while still enforcing the relationship and constraints between the values. However, programmers often find polymorphic types difficult to use and understand and tend to reason using concrete types. We propose GeckoGraph, a graphical notation for types. GeckoGraph aims to accompany traditional text-based type notation and to make reading, understanding, and comparing types easier. We conducted a large-scale human study using GeckoGraph compared to text-based type notation. To our knowledge, this is the largest controlled user study on functional programming ever conducted. The results of the study show that GeckoGraph helps improve programmers' ability to succeed in the programming tasks we designed, especially for novice programmers.
\end{abstract}

\maketitle

\section{Introduction} \label{sec:intro}
In programming languages, a polymorphic type \cite{Cardelli1987-fp} can represent values of different types while providing a common interface or behavior for those values. Polymorphic types are central to the succinctness of contemporary statically-typed functional languages, enabling a considerable degree of type-safe abstraction and hence component reuse. Parametric polymorphism is available in many programming languages, from functional languages such as Haskell and ML to imperative and multi-paradigm languages such as Rust\cite{Klabnik_undated-wx} and Go\cite{Griesemer_undated-ff}. Polymorphism allows programs to be written in a way that is more generic and adaptable to different data types, enabling greater flexibility and code reuse. Polymorphic types are ideal for modeling abstractions, such as properties of mathematical objects and laws that hold on these objects. 

\begin{figure}[hbt]
  \includegraphics[width=\linewidth]{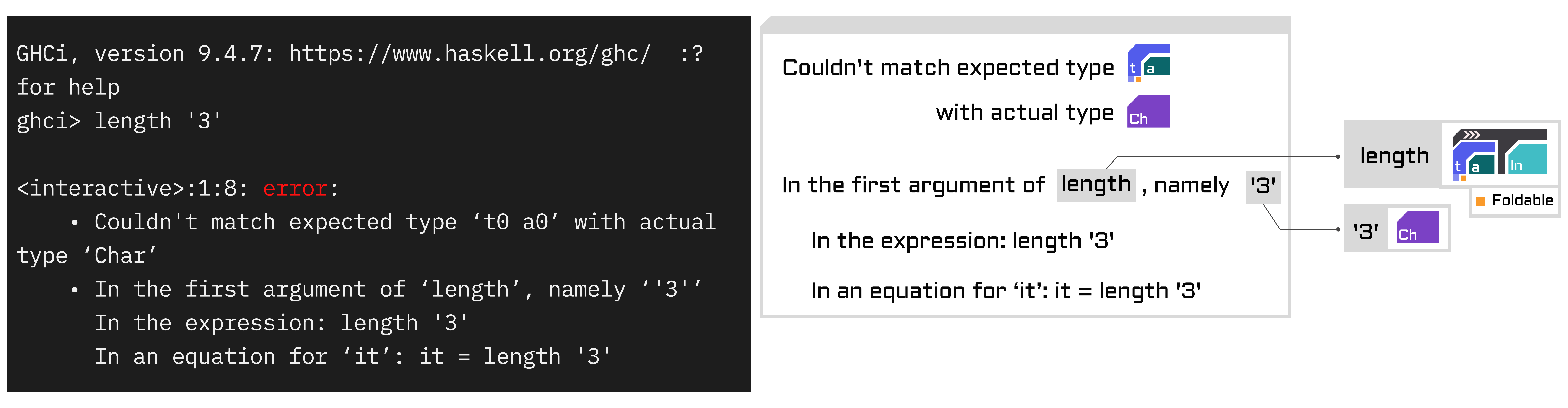}
  \caption{\label{fig:example-foldable} An example type error where a programmer mistakenly provided a \texttt{Char} instead of a \texttt{String} literal. \textbf{Left --} The compiler shows an error message comparing the provided Char type to a confusingly named type \texttt{t0 a0}. \textbf{Right --} GeckoGraph shows the exact message with the two types in graphic notation, highlighting the structural difference rather than identifier names.}
\end{figure}

Although polymorphic typing promises robustness and a high degree of code reusability, studies~\cite{Jun2000-ec, Jun2000-yu} show that using polymorphism in practice poses challenges, especially for novice users. These studies have shown that humans tend to focus on concrete types and only rely on polymorphic type checking as a last resort. In practice, polymorphic types often pose usability problems for programmers. New polymorphic type variables can be created during type checking. These intermediate type variables are often kept behind the curtain unless a type error is encountered. This often results in programmers resolving type errors with type variables not authored by the programmers themselves (Fig. \ref{fig:example-foldable}).

While type polymorphism is one of the oldest topics in programming language theory \cite{Cardelli1987-fp}, little research focuses on the \textit{usability} of polymorphic types. Hage argues that the expressiveness and power of type systems often come at the cost of usability~\cite{Hage2020-hg}. We aim to investigate the challenges of using polymorphic types and explore how to improve their usability with visualization and modern HCI techniques. To achieve this, we propose GeckoGraph, a graphical notation for types. GeckoGraph aims to complement traditional text-based type notation and make reading, understanding, and comparing types easier. GeckoGraph is prototyped and verified iteratively, leading to a design with visual clarity applicable to many programming contexts. Our study evaluating 714 participants' ability to solve type adaptation challenges in GeckoGraph versus text-based type annotation is, to our knowledge, the largest controlled user study of a functional programming tool or concept. The study results show that GeckoGraph helps improve programmers' ability to succeed in resolving the type challenges, especially for novice programmers.

\section{GeckoGraph}

GeckoGraph is a visual notation for type annotations in statically typed programming languages. It is intended to work tangibly with text-based annotations, but uses colors, shapes, and symbols to make structures of types easy to identify at a glance. In this section, we describe the design of GeckoGraph and highlight some unique benefits of programming with GeckoGraph.

\subsection{Design of GeckoGraph}
The design of GeckoGraph focuses on visualizing types in functional languages (e.g., Haskell, ML). In this paper, we use Haskell as an example. As illustrated in this section, it can express basic types, polymorphic types, algebraic data types, and some advanced type-level features. However, GeckoGraph could also be used in imperative and multiparadigm languages such as Rust. We provide a GeckoGraph construction library for Haskell. 
        
We identified three main design goals for GeckoGraph based on the challenges of using polymorphic types~\cite{Jun2000-ec, Jun2000-yu} and how programmers tend to use type annotations~\cite{Justin_Lubin2021-yy}, as follows. 

\paragraph{\textbf{(D1) Low barrier to learn}}\label{goal1} GeckoGraph should take little to no effort to learn. The rules to translate a text-based type notation to GeckoGraph should be minimal. Where possible, GeckoGraph should be intuitive to programmers who are familiar with text-based type notation.

\paragraph{\textbf{(D2) Easy to parse for humans}}  \label{goal2} GeckoGraph should make the task of reading and understanding type notation easy. It should emphasize the less obvious properties of a type signature. GeckoGraph should eliminate the need for mental backtracking, such as counting opening and closing parentheses and remembering which type classes are required on which variables.

\paragraph{\textbf{(D3) Easy to compare and search}} \label{goal3} GeckoGraph should aim to make the task of comparing two types easy, especially to make subtle differences in text-based notation harder to miss. This also includes the task of choosing an ideal function from a list of potential functions. For example, programmers search for a desired function from a documentation site with only partial knowledge of its type (e.g., the arity, one of the argument types, or type class it must fulfill).

\paragraph{Simple Types} 
Simple types, such as type variables and concrete types, are displayed in a cell \includegraphics[height=1em]{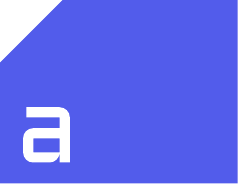}: a solid-colored rectangular box with an angled corner on the top left. Each type identifier encodes a distinct color hue of the cell.  Its first 1 or 2 letters are displayed inside the cell at the bottom left to provide familiarity (design goal \ref{goal1}) and strong secondary encoding. The angled corner in the top left provides visual separation between two cells, even when the same color cells are next to each other, allowing GeckoGraph to be zoomed out to extremely small sizes (Section \ref{subsec:space}) without suffering readability (design goal \ref{goal2}).

\begin{figure}[hbt]
  \includegraphics[width=\linewidth]{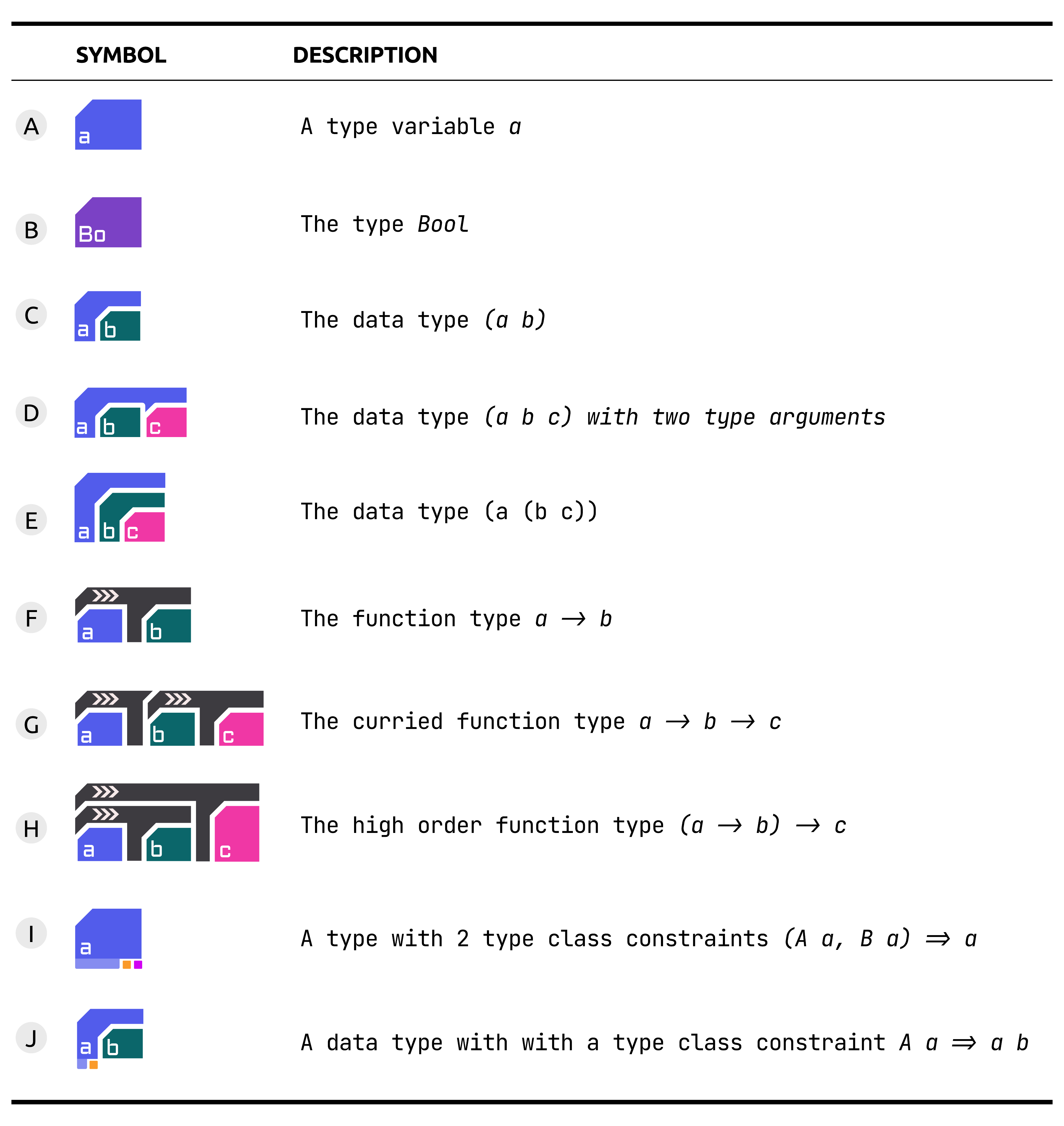}
  \caption{
        \label{fig:design}
        Examples of various types as represented in GeckoGraph, including type variables (A) and concrete types (B). Data types (C, D, E), function types (F, G, H), and type classes (I, J).
  }
\end{figure}

\paragraph{Data types}
GeckoGraph displays an algebraic data type as a larger cell, where the type constructor half encloses its arguments: \includegraphics[height=1em]{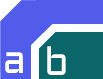}. The arguments are aligned in the bottom right of the cell. Two distinct visual dimensions are used to provide additional visual clarity (design goal \ref{goal2}). Data types containing more arguments (e.g., \texttt{ a b c} or \texttt{(a b) c}) will expand horizontally \includegraphics[height=1em]{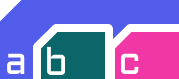}. Data types that are nested (e.g. \texttt{ a (b c)}) will grow taller \includegraphics[height=1.2em]{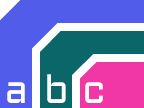}. This distinction accommodates our design goal \ref{goal3}. Note that the height of GeckoGraph grows only upwards, but not downwards. Not only does this allow GeckoGraph to be more efficient in its space usage, but it also allows the legend text to be correctly aligned at the bottom and can be read similarly as regular type notation (design goal \ref{goal1}).

\paragraph{Function Types}
Functions are the fundamental building blocks of functional programming languages, and function types are ubiquitous and the most important in type-level programming. In Haskell, \texttt{(->)} is defined as an infix type operator with the right associativity to provide succinct type annotation. GeckoGraph preserves this syntax feature to make the notation more intuitive (design goal \ref{goal1}): the 2 arguments of a function type in GeckoGraph are placed on both sides of the cell \includegraphics[height=1em]{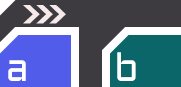}. A special function indicator (\texttt{$>>>$}) is displayed at the top of the cell.

\begin{figure}[hbt]
  \includegraphics[width=\linewidth]{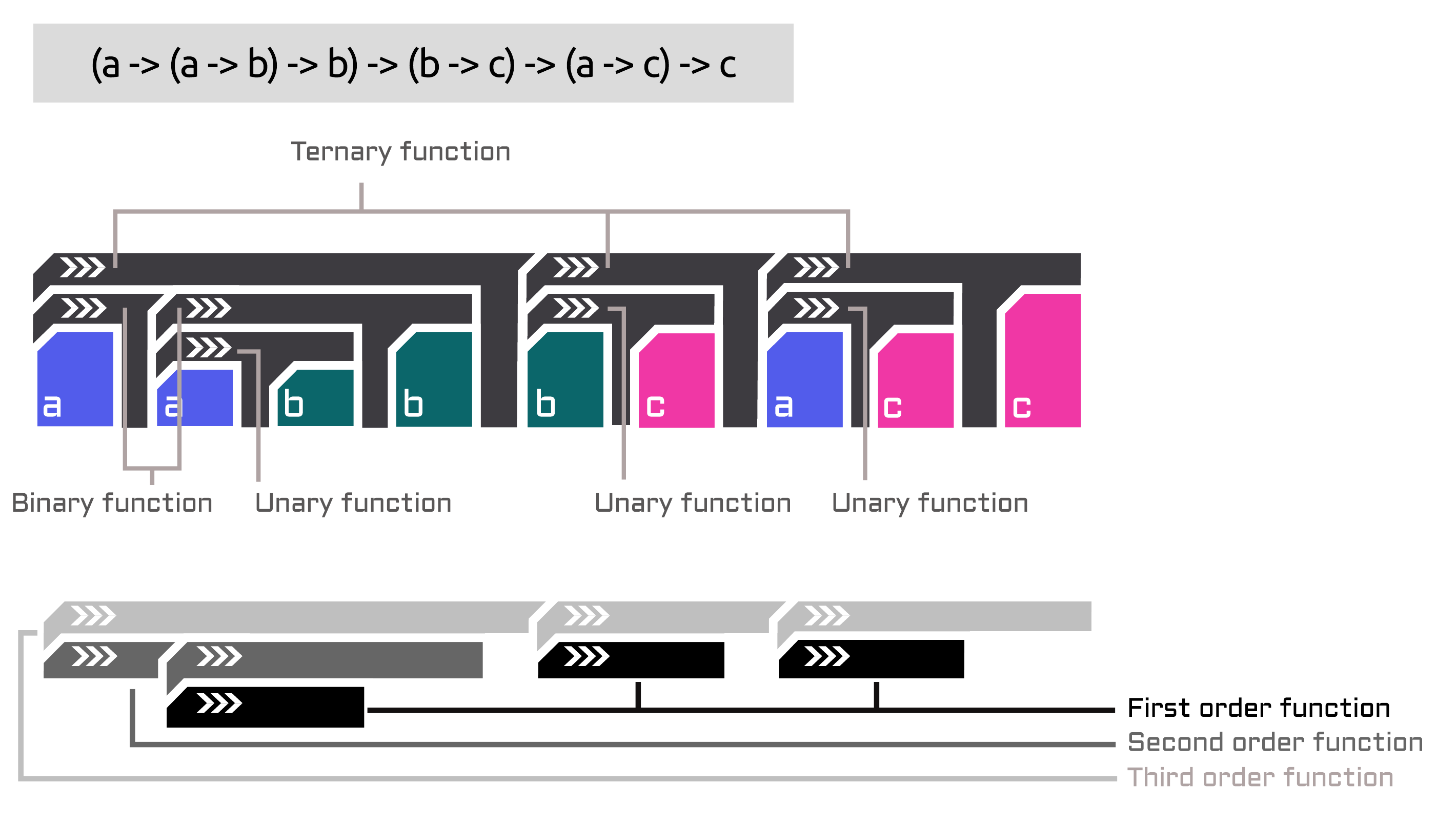}
  \caption{
        \label{fig:indicator}
        An example of using the function indicator. The function indicator can be used to easily identify the arity of a function type by counting the connecting function indicators. For high-order functions where functions are arguments of other functions, it is very easy to see the ``order" of functions and how they are arranged. 
  }
\end{figure}

Curried functions (e.g., \texttt{ a -> b -> c}) display as two cells of functions merged together \includegraphics[height=1em]{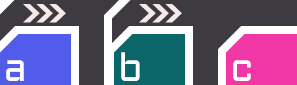}; the second function overlaps on top of the first, indicating that the second function is the return type of the first. Regular high-order functions (e.g., \texttt{(a -> b) -> c} ) follow the rules of functions and nested data types \includegraphics[height=1.2em]{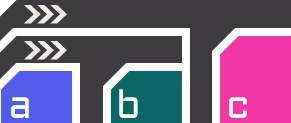}. The placement of function indicators aims to make it easy to find desired functions in the documentation site based on function arity and high-order functions (design goal \ref{goal3}). It is easy to tell high-order functions from the vertical position of its function indicator. Similarly, it is easy to count the arity of a function by counting the number of horizontally connected function indicators (Fig. \ref{fig:indicator}). 

\paragraph{Type Classes} 
Type classes are an intrinsic part of Haskell \cite{Hudak2007-kn}, and many other functional languages. In GeckoGraph, the type classes (e.g., \texttt{ (A a, B a) => a}) are indicated in the extended area below one or more GeckoGraph cells \includegraphics[height=1.2em]{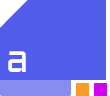}. Each type class required on a type variable displays as a square indicator aligned on the right of the extended area. In GeckoGraph, type-class constraints are associated with every instance of the type variable that requires them. This means when displaying the type \texttt{(==) :: Eq a => a -> a -> Bool} in GeckoGraph, the constraint \texttt{Eq} appears in both occurrences of \texttt{a}. 

The GeckoGraph type class's design promotes the type class placements rather than the type class names. Programmers can easily see where and how many type classes are required, but they may need an extra step (global legends of the color mapping or pop-up window) to identify the name of the type class. We believe that this decision is well justified. For example, when reading a type \texttt{(A a, A c, B a, B b, C b) => a -> b -> c}, programmers may need to switch back and forth to remember which type classes are needed on which variable. GeckoGraph helps minimize the effort to associate each type variable with all its type classes (design goal \ref{goal2}).

\subsection{Benefits of Using GeckoGraph}\label{sec:benefits}
\paragraph{Generalization patterns in type classes}
A frequently cited confusion among novices is the overlap between list type \texttt{[a]} and its more generic counterpart \texttt{Foldable} instances. The subtle differences are often not important for beginners. However, when encountering type errors in working with lists, Haskell often explains the error with the \texttt{Foldable} instance. For the example in Fig. \ref{fig:example-foldable} where a programmer mistakenly provided a \texttt{Char} instead of a \texttt{String} literal, the compiler shows an error message comparing the provided \texttt{Char} type with a confusingly named type \texttt{t0 a0}. Although any \texttt{Foldable} instance is perfectly suited for the list \cite{Waldmann2018-hu}, this generalization may reduce programmers' confidence in their understanding of the language and the ability to navigate out of a type error. In GeckoGraph, a list type and a type with a Foldable instance have the same shape. This allows the generic type \texttt{t0 a0} in the error message to assume the same shape as \texttt{[a]}. This generalization of concrete types and abstraction of type-class instances aims to allow for teaching fundamental functional programming concepts without hiding high-level abstractions. The same benefits apply to polymorphic numbers and strings.

\paragraph{Consistent color scheme}\label{par:color-scheme}
A common task in programming is to scan for a desired function from a sea of potentially useful functions, such as library documentation. During scanning, programmers often have partial knowledge of the desired function, e.g., the arity, one of the argument types, or the type class it must fulfill.  A typical example is conversion: using a known \texttt{String} type to produce a desired \texttt{Data.Text}  type. Another example is the `lookup' function: using a known \texttt{Data.Map a b} to produce a desired \texttt{b} type. GeckoGraph supports this task by using consistent colors for the same type identifier. Programmers can rely on the color grouping to scan for the desired type in their project or in third-party library documentation.

\paragraph{Advanced Type Feature Visualization}
The design of GeckoGraph enables the visualization of many advanced type-level features. \textbf{Kind visualization}: if the kind of type variables can be inferred, the kind information is consistently displayed in GeckoGraph. For example, in figure \ref{fig:advanced} (A), the variable \texttt{a} needs at least the kind \texttt{* -> *} because of its use on the right-hand side. GeckoGraph respects this kind information and displays it as a constructor type over an empty structure, indicated using a dotted outline.  \textbf{Qualified constraints}: GeckoGraph's type class notation naturally extends to support qualified constraints. In the type \texttt{forall b. A (a b) => a b}, GeckoGraph shows the scope type class requirement on \texttt{a b} (Fig. \ref{fig:advanced} B).
\textbf{Multiple Parameter Type Class}:  GeckoGraph supports multiple parameter type classes by using multiple shapes with the same color hue to indicate the different parameters of the same type class.  For example, for the type \texttt{A a b => a b},  GeckoGraph shows that the variables a and b both need an A class, but they are the different parameters of A (Fig. \ref{fig:advanced} C).

\paragraph{Precise Interactivity}
Modern programming environments often allow programmers to mouse over part of the source code to query detailed information, such as definition, references or documentation. However, with text-based source code, it is often hard to distinguish whether programmers want the most specific fragment under the cursor or larger blocks. Because of its graphical layout, GeckoGraph allows programmers to precisely select which part of a type signature they intend to query, that is, in Fig. \ref{fig:advanced} (D) when the user mouses over the type class box (orange square) under the second occurrence of \texttt{a} the type class it represents is revealed in detail. 

\begin{figure}[hbt]
\label{fig:advanced}
  \includegraphics[width=\linewidth]{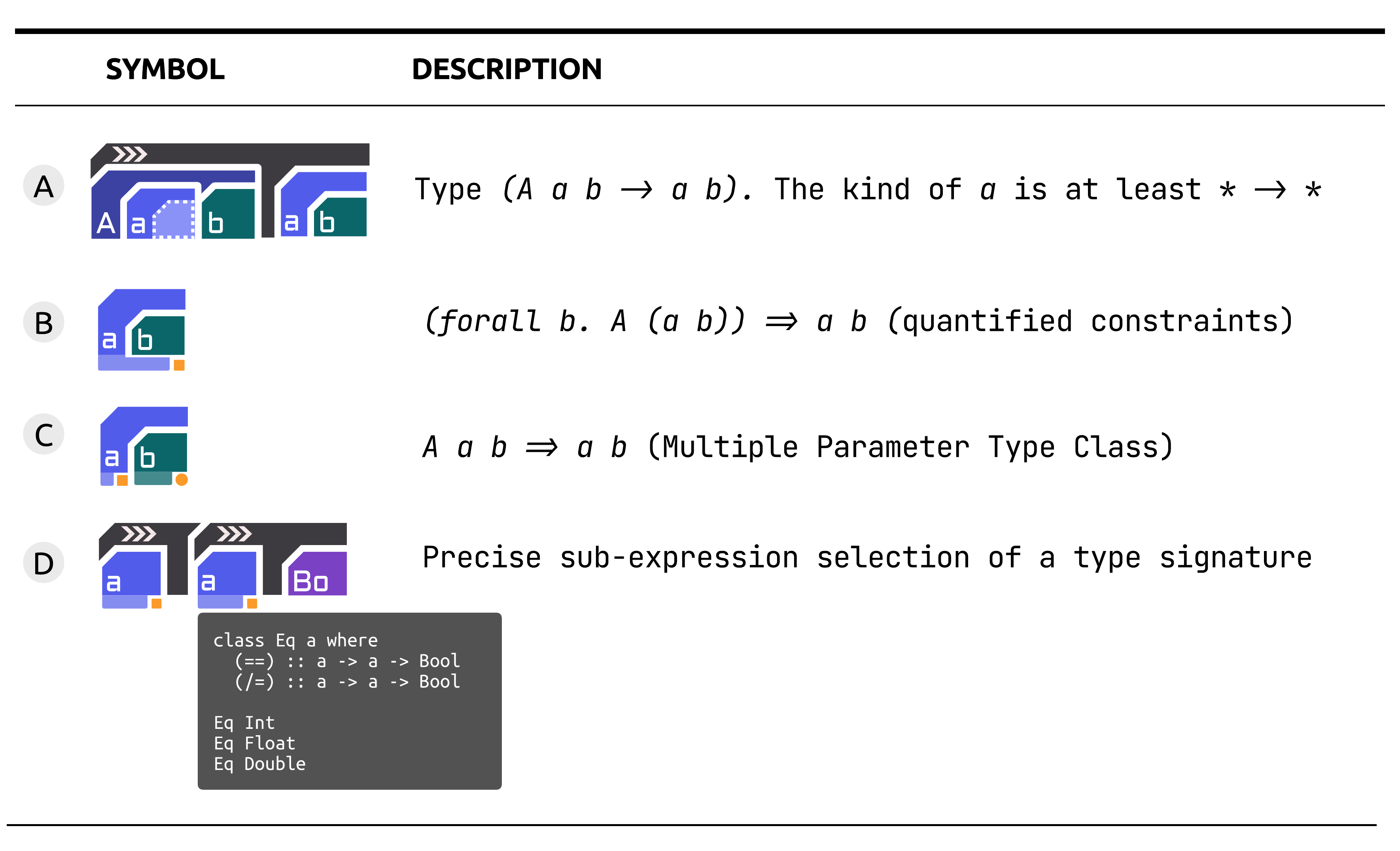}
  \caption{
  \label{fig:advanced}
  Advanced features of GeckoGrap. (A) GeckoGraph supports Kind Visualization if the inferred kind is greater than \texttt{*}. (B) GeckoGraph supports qualified constraints by extending the extended area across multiple type variables. (C) GeckoGraph supports Multiple Parameter Type Classes, using different shapes of the same color to indicate that multiple variables must satisfy certain type classes collectively. (D) GeckoGraph supports the precise selection of its sub-structures. }
\end{figure}

\paragraph{Language Agnostic}
GeckoGraph can be implemented in any language that uses static typing. In programming projects, GeckoGraph supports polyglot programming projects. Typical circumstances include projects using foreign function interfaces or multiple languages for client- and server-side programming. GeckoGraph provides a common notation to describe the functionality and features of systems. In teaching and learning programming languages, GeckoGraph removes the nomenclature difference in different programming languages.  For example, when describing algebraic data types, different language communities use various names: tuple, enum, struct, etc. It is important to realize that these are the same concepts and ignore the minute linguistic barriers.

\subsection{Previous iterations of GeckoGraph}
GeckoGraph was designed through many different iterations. Many research methods were used to verify ideas, including prototyping, cognitive walk-throughs, and formative studies.  We list some notable design elements and their major feedback.

\begin{figure}[hbt]
  \includegraphics[width=\linewidth]{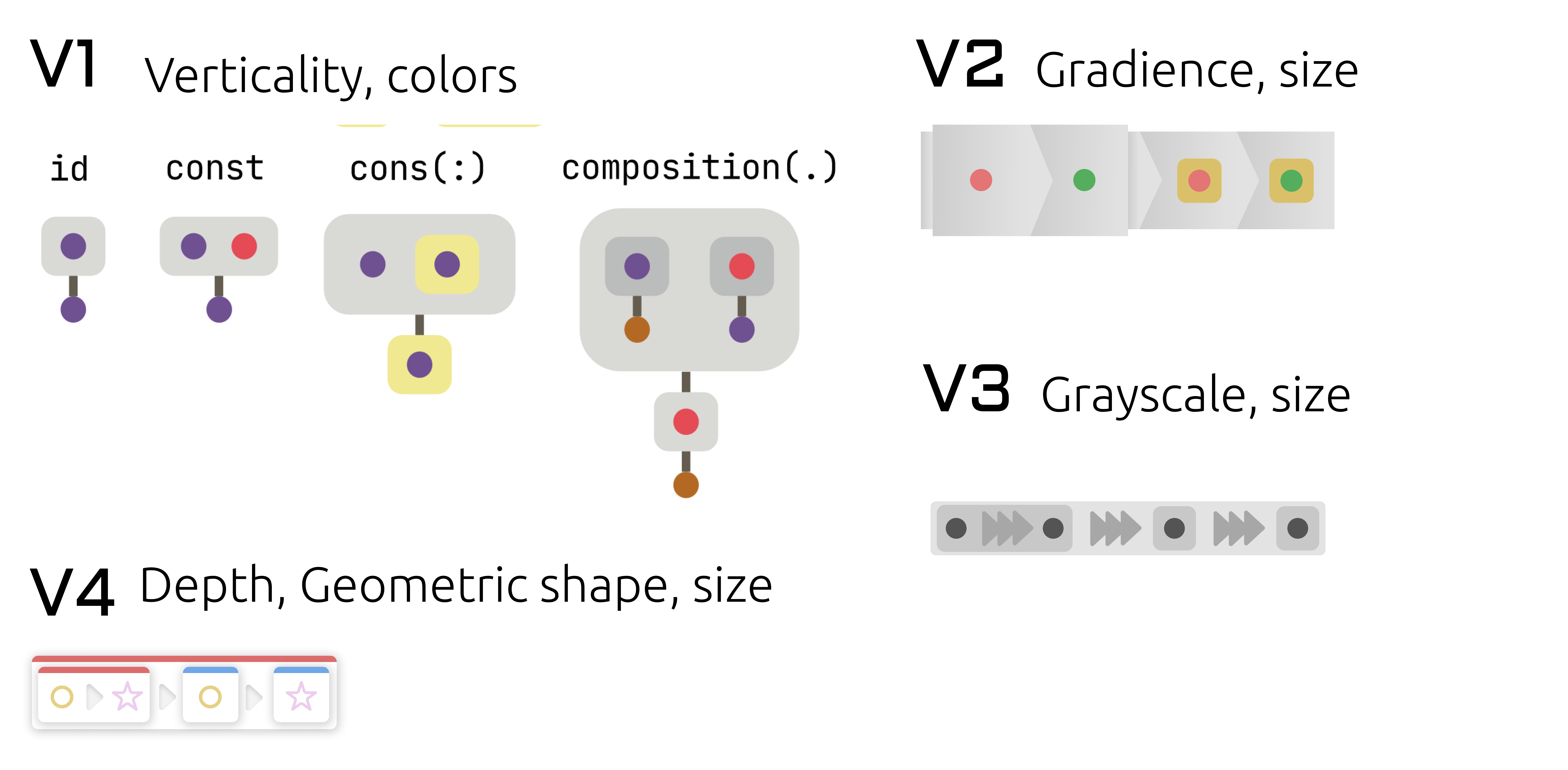}
  \caption{\label{fig:previous}Previous Versions of GeckoGraph. Different encodings represent named types, type variables, type constructors, and high-order functions.}
\end{figure}

\paragraph{\textbf{Encoding the direction of function types}}
This can be found in the design v1 (Fig. \ref{fig:previous}). This design shares many similarities with a prior project \cite{Jung2000-oc} in type visualization. Although many students and experts agree on this, its use of space on both width and height scales is proportional to the types in question, causing too much inconvenience. In addition, this design does not produce a canonical form for a curried function. For example, for composition (.), it is unclear whether it takes two functions as input and returns a binary function or two functions and a single structure as input and returns a different structure.

\paragraph{Encoding the depth and size} Grayscales, gradience, size, and simulated 3-dimensional elevation are promising visual representations for depicting numeric dimensions such as the depth and width of the parse tree. However, gradient and elevation were dismissed because of their requirements for a more demanding rendering process, making GeckoGraph harder to implement in more restricted user interfaces, such as the command line. In addition, all of these visual dimensions reduce readability when scaled down to a very small size.

\paragraph{Encoding symbolic names} In some variations, we tested using different geometric shapes to indicate the symbolic names of type variables and concrete types. We decided against using icons due to the limited number of different shapes until they were indistinguishable. The color provides more encoding spaces, and the letters provide familiarity with the original type annotations. This was shown to help reduce friction in the adoption of GeckoGraph.

\section{Evaluation}
To evaluate the usefulness of GeckoGraph, we designed a controlled experiment in the form of a game called "Zero to Hero". The game contains 10 levels of varying difficulty. At each level, participants are asked to implement a function called "\texttt{zeroToHero}" using only a list of available functions. These available functions are different at each level, and the target types of Zero and Hero vary at each level. The details of each level are provided in the Appendix (Appendix \ref{levels}). 

The experiment aims to study how polymorphic types are used and reasoned about during programming tasks. In particular, we studied how programmers scan and select potentially useful functions from a library and compare intended types and actual types during type errors.

\subsection{An example level}
\begin{figure}[hbt]
  \includegraphics[width=\linewidth]{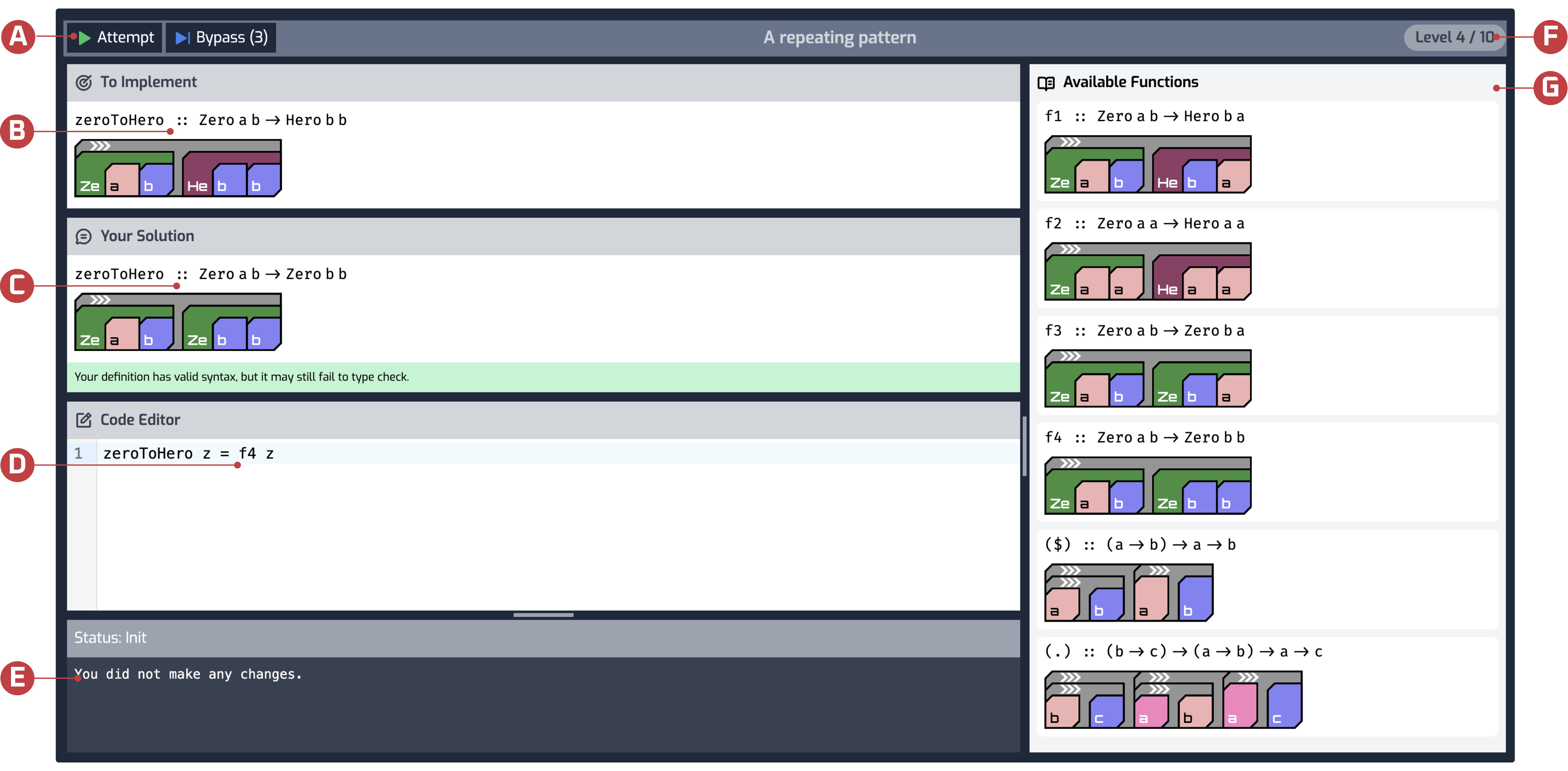}
  \caption{\label{fig:level-example} A screenshot from  the game ZeroToHero. On this level -- level 4 (Shown at F) -- the players need to implement the function \texttt{ zeroToHero :: Zero a b -> Hero b b} (B). They write their own definitions in the code editor (D) using a set of provided functions (G). The inferred type of their current definition is shown in (C). When ready, they can test their solution by clicking on the \textit{Attempt} button (A). They can also skip a level by clicking on the \textit{Bypass} button next to it. The output from the compiler, if there is any, is shown in a window below (E). The GeckoGraph in the screenshot uses a different color scheme that is optimized for computer screens.}
\end{figure}

We illustrate the task of the user study using level 4 of the game. At this level (Fig. \ref{fig:level-example}), the programmers are tasked with implementing the function \texttt{ zeroToHero :: Zero a b -> Hero b b}. The available functions are \texttt{f1::Zero a b -> Hero b a}, \texttt{f2::Zero a a -> Hero a a}, \texttt{f3::Zero a b -> Hero b a}, and \texttt{f4::Zero a b -> Hero b b}. Two generic functions \texttt{(\$)} and \texttt{(.)} are provided to improve the ergonomics of composing functions, but all tasks can be won without the use of generic functions. The possible solution and other details of the level can be found in the appendix (Appendix \ref{levels}).

At each level of the game, the programmer must select the right functions to achieve the target result. In particular, for this level, participants must discover that only \texttt{f4} and \texttt{f2} are necessary to produce the desired results. An implement that satisfies the target type is \texttt{zeroTohero z = f2  (f4  z)}.

\subsection{Recruitment}
Participants were recruited online through the Haskell community on Reddit and Discord. Participation is fully anonymized; detailed ethical implications of these experiments were reviewed and approved by the IRB of the authors' institution.

\subsection{Group Assignments}

The experiment uses a between-subject design. However, all participants receive both treatments (with and without GeckoGraph) during their runs.
Participants are assigned to one of two groups. Both groups receive the same tasks in the same order. Group one participants are assisted by GeckoGraph on even levels and only text-based type annotation on odd levels. Group two participants are the same but with the order flipped. Both groups have access to the text-based type signature for all tasks. The number of participants in the two groups is counterbalanced.

\subsection{Hypothesis}
In programming tasks that involve reading and understanding polymorphic types, graphic notation using visual elements that provide high grouping strength (colors, shapes, sizes, and symbols) can improve the performance of such tasks compared to traditional text-based type notation. Our null hypothesis is that "Using graphic notation has no effect compared to traditional text-based type notation." This hypothesis and the task design were registered at the Open Science Foundation prior to data collection. 

\subsection{Task Design} \label{subsection:task}
Participants in both groups receive the same 10 tasks. The tasks start off easy but gradually increase in difficulty.  In each task, a target type signature of the function \texttt{zeroToHero} is given to the participants. Participants are provided with a list of available functions to implement the target function. This is to simulate the tasks of selecting useful functions from a library. In addition, participants are not allowed to use any other functions or variables outside the provided functions; even the Haskell prelude is not available. This ensures that everyone has the same knowledge and minimizes the effect of familiarity. 

Participants can skip a level during the game if they are stuck. We believe that it is normal for anyone to get stuck on a challenging task, and being stuck on one of the 10 tasks does not discount their qualitative input of the tool. We limit the number of skips that a participant can use during the game to four times so that submitting qualitative feedback without completing at least some levels is impossible. 

\subsection{Measurements}
During the study, the time spent by participants on each task is recorded. We also record the resulting status of each level, whether it is a success or failure. Before each run, participants nominate their level of Haskell experience on a four-level scale: beginner, familiar,  knowledgeable, and expert.  If a participant has completed all 10 levels (with the help of skipping), we invite the participant to complete a post-study survey. In it, we ask for their opinion on how intuitive the GeckoGraph design is, how distracting they find GeckoGraph, and how helpful GeckoGraph is during the game, using a seven-point scale. In the end, we ask a few open-ended questions, inviting participants to provide their experience using GeckoGraph and their expectations about the potential applications of GeckoGraph.

Data collection from the human study was stopped after the planned cut-off period of 14 days. After the cut-off date, the ZeroToHero game is open source and available for free evaluation \cite{Anonymous_undated-ne} and repeating our experiment, but no further data was collected. 

\section{Results}

During the data collection period, a total of 714 users participated in the study. Among them, 245 are novice users, 216 are familiar with Haskell, 216 are knowledgeable users, and 88 are expert users.

\subsection{Time to complete levels}

\begin{figure}[hbt]
  \includegraphics[width=\linewidth]{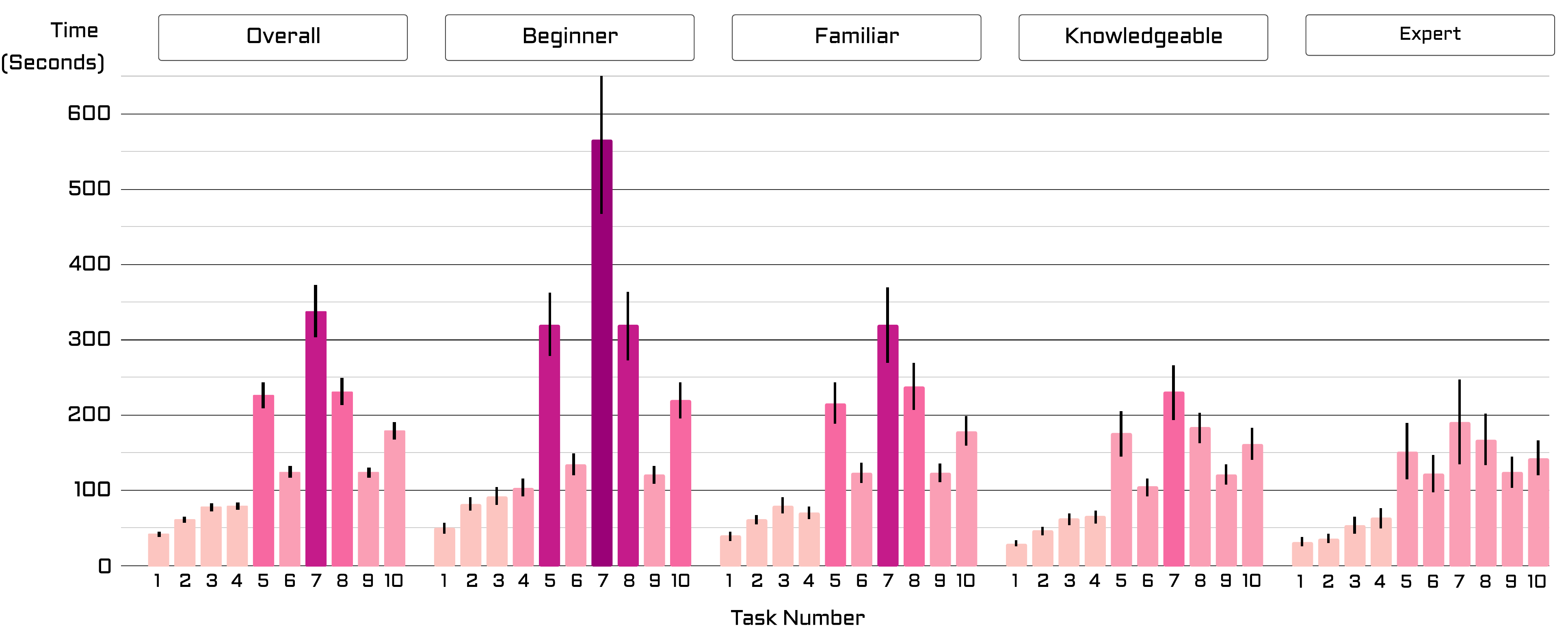}
  \caption{\label{fig:level-time} Time spent on each level, with 95\% confidence interval. We show that the difficulty steadily increases across the game, but levels 5, 7, and 8 are significantly harder than the authors intended. The overall task time of each group roughly matches experience level.}
\end{figure}

The 10 levels are designed to gradually increase difficulty. From the results of the experiment, most of the tasks align with this trend. However, three tasks stand out in Fig. \ref{fig:level-time}.  Level 7 (mean = 334 seconds) is the hardest task in the game in terms of time, followed by level 8 (mean = 228 seconds) and level 5 (mean = 224 seconds). To complete an average level, the beginner group uses an average of 100 seconds, the familiar group uses 90 seconds, the knowledgeable group uses 80 seconds, and the expert group uses 70 seconds. This roughly aligns with self-reported expertise. We show that the task time on each level follows normal distributions using a Shapiro-Wilk test \cite{Shaphiro1965-dx} (p-value  $ \leq 1.018 \times 10^-16$, for an alpha value of 0.05, p less than 0.05 is considered normal distribution).

Level 5, 7 and 8 are the only three levels that include functions from standard Haskell library, baring the \texttt{(.)} and \texttt{(\$)} provided for convenience. Level 3 requires programmers to use the \texttt{fst} and \texttt{snd} functions to extract value from a tuple. Level 7 requires programmers to use the \texttt{(<*>)} function of the \texttt{Applicative} class, while level 8 the \texttt{fmap} function of the \texttt{Functor} class. The authors speculate that the more experienced participants are much more familiar with these functions, hence the strong contrast on these three levels.

However, when comparing the task time between the two treatments, we were unable to reject the null hypothesis. In a two-sample T-test, we could not find any significant difference between the two groups overall (p-value = 0.457), nor does there differ between the two groups in any of the four levels of experience (beginner: p-value = 0.845, familiar: p-value = 0.524, Knowledgeable: p-value = 0.712, expert p-value = 0.771).

\subsection{Success rate}
We saw that, overall, GeckoGraph provides a higher success rate (96.88\%) than text-based type notation (94.62\%). This trend can be seen in every experienced group: beginner group (95. 12\% vs. 92. 68\%), familiar group (97. 39\% vs. 93. 34\%), knowledgeable group (96. 82\% vs. 96. 06\%) and expert group (98. 2\% vs. 96. 40\%). We saw the significance decrease as the user's experience increased. When performing a proportion test on each group, we see that the effect is most significant with the beginner group and reject the null hypothesis (z score = 2.0228, p-value = 0.0431), followed by the familiar group (z score = 1.7495, p-value = 0.0802). The knowledgeable group (z score = 1.0295, p-value = 0.3032) and the expert group (z score = 0.8660, p-value = 0.3756) show less significant differences between treatments. 

\begin{figure}[hbt]
  \includegraphics[width=\linewidth]{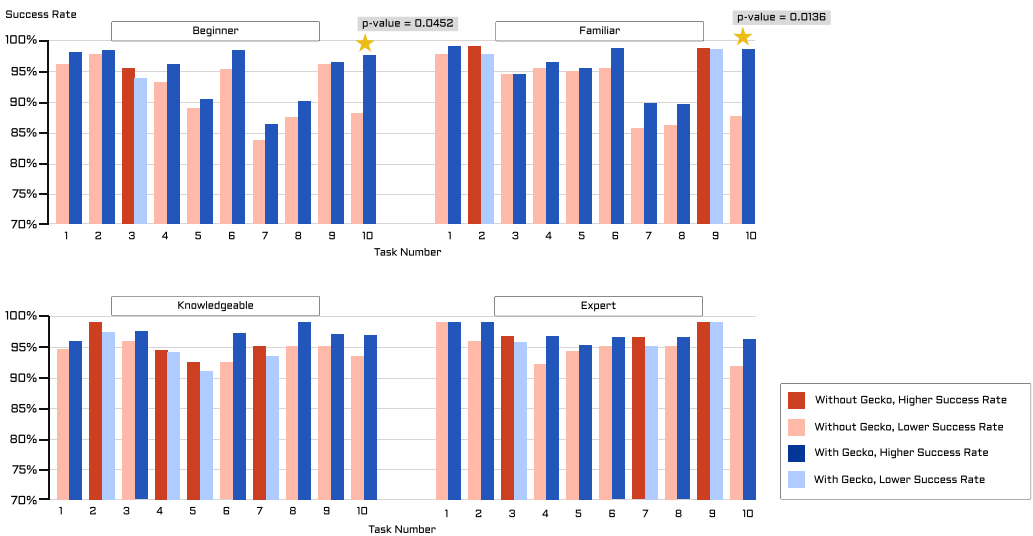}
  \caption{\label{fig:success-rate}Success rate of each task with and without GeckoGraph, grouped by experience level. The figure is cropped from 70\% to 100\% for readability. In most tasks, GeckoGraph provides a small edge. However, significant differences were found in task 10 of the beginner group and task 10 of the familiar group. }
\end{figure}

When breaking down the result in each task (Fig. \ref{fig:success-rate}), we were able to reject the null hypothesis in task 10 of the beginner group and task 10 of the familiar group \ref{fig:success-rate}. We will address this correlation in Section~\ref{sec:discussion}.

\subsection{Qualitative Feedback}
\begin{figure}[hbt]
  \includegraphics[width=\linewidth]{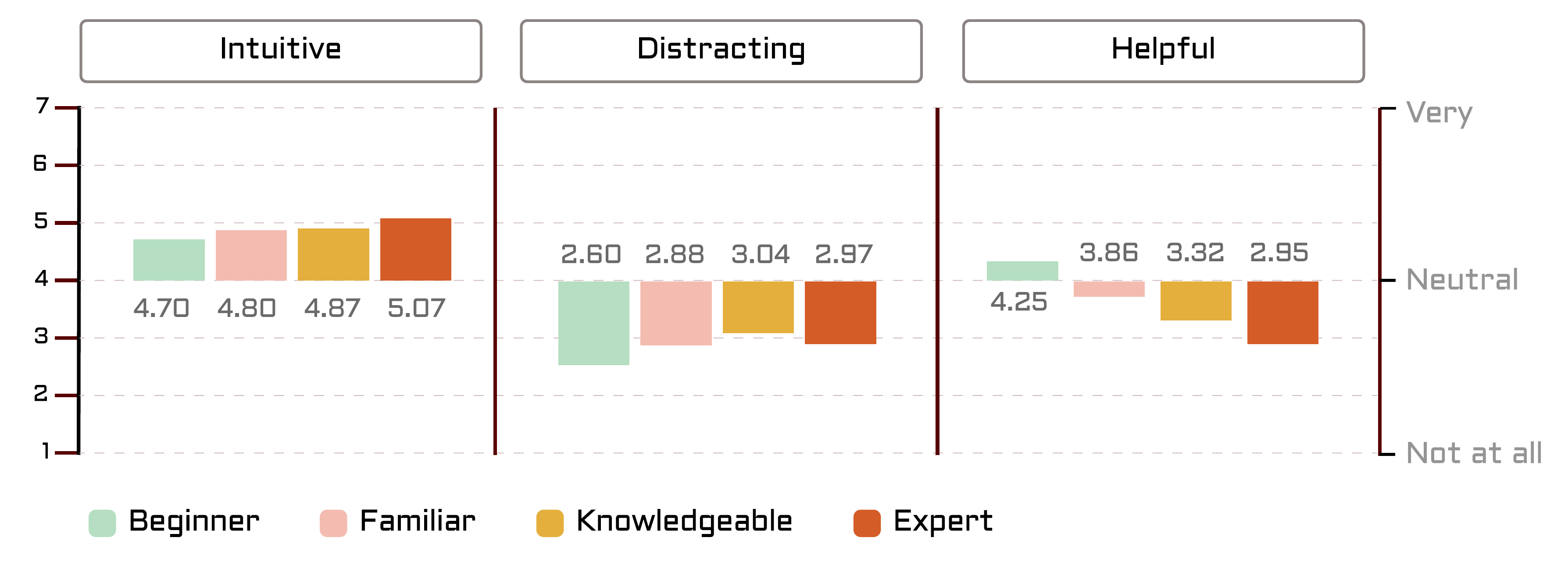}
  \caption{\label{fig:qualitative} The users rated scores of how intuitive (left), distracting (middle), and helpful (right) they found GeckoGraph. Overall, programmers consider GeckoGraph to be intuitive and not distracting. However, opinions are split on its helpfulness. }
\end{figure}

In their responses to the post-study survey, most programmers believe that the design of GeckoGraph is intuitive and that its appearance in the interface does not cause distraction.
For the question ``Do you find the GeckoGraph distracting", most of the participants rated a negative score, with an average of 2.88 (Fig. \ref{fig:qualitative} left). For the question ``How intuitive do you find the GeckoGraph?", we saw a reverse correlation of experience (Fig. \ref{fig:qualitative} Middle): experts find the GeckoGraph most intuitive (5.07), followed by the knowledgeable group (4.87), and the familiar group (4.80). The beginner group found it to be the least intuitive but still rated a positive score of (4.71). 
 When answering the question ``How helpful do you find GeckoGraph in finding the solution during the game?", the answer is more divided into different experience groups (Fig. \ref{fig:qualitative} right). It is slightly positive for beginners (4.25) and slightly negative for the other groups, the familiar group (3.86) and the knowledgeable group (3.32). The expert group finds that GeckoGraph is relatively unhelpful, with an average score of 2.95.

\subsection{Threats to validity}

\paragraph{Task design}
In our human study, most of the provided functions are very abstract. These functions are created by the authors solely for the gamified study. They are designed to be different from well-known Haskell functions to minimize the familiarity variable. They are also designed with an interest of being puzzling and fun.  These functions may not be the most representative of real-world Haskell programming.

\paragraph{The use of skips}
Although we justified the use of skips in Section \ref{subsection:task}, the availability of skipping does allow users to adopt more utilitarian strategies, often involving skipping a level without giving it a fair try. This happened more often in the later levels when users realized they had enough skip opportunities left to ``complete the game". These strategies may result in lower recorded success rates than if no skips were allowed.

\section{Discussion} \label{sec:discussion}
\subsection{Strengths}
From the results of our experiment, we see that using GeckoGraph has a significant effect on the success rate of our participants, especially on less experienced programmers. We also see that with the data we collected, we did not find a significant time difference between programming with and without GeckoGraph. To extrapolate the observed expressiveness, we speculate on the practical benefits of programming with GeckoGraph.

\subsubsection{Identify the Most Important Features}
One trend that we saw from the qualitative feedback is that programmers find GeckoGraph helpful for finding patterns and important features of the types. Programmers are very positive about GeckoGraph's ability to reveal the most helpful features of a type in distinctive visual elements such as color, length, and height.

\begin{figure}[hbt]
  \includegraphics[width=0.6\linewidth]{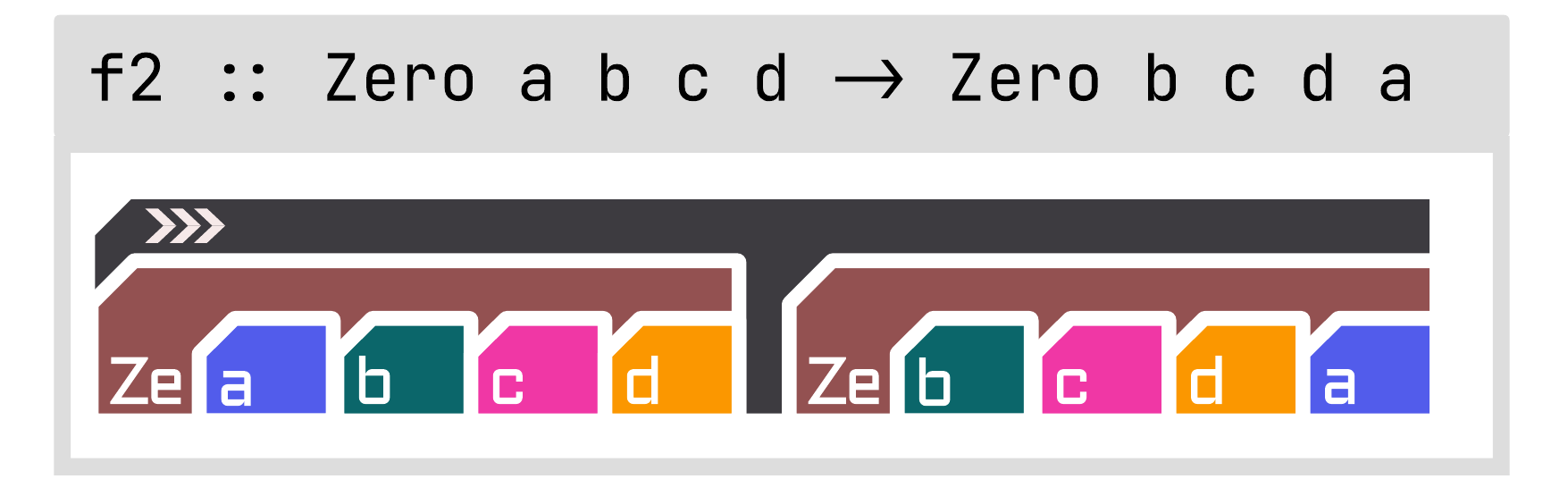}
  \caption{\label{fig:rotate} The `rotate' function in level 8 of the user study. The name given in the game is `f2'. It shuffles the type arguments of a Zero type}
\end{figure}

\textbf{The colors of GeckoGraph} help programmers to see the permutation of type variables in the input and output of a function. A recent review \cite{Zeng2023-jz} of 59 graphical perception articles showed that combining solid color hue in a filled shape provides stronger visual perception for nominal data such as type identifiers. One example of GeckoGraph's effective use of color is the ``rotation" function in the user study (Fig. \ref{fig:rotate}). With text-based type notation, programmers often rely on mnemonic devices such as alphabetic ordering or naming conventions. For example, the rotation function \texttt{f2 :: Zero a b c d -> Zero b c d a} in the game is less recognizable if changed to \texttt{f2 :: Zero e v m h -> Zero v m h e}. To quote a participant, ``GeckoGraph is quite intuitive to see the permutations of the arguments. Also, to see how to produce and consume arguments." 

\begin{figure}[hbt]
  \includegraphics[width=0.6\linewidth]{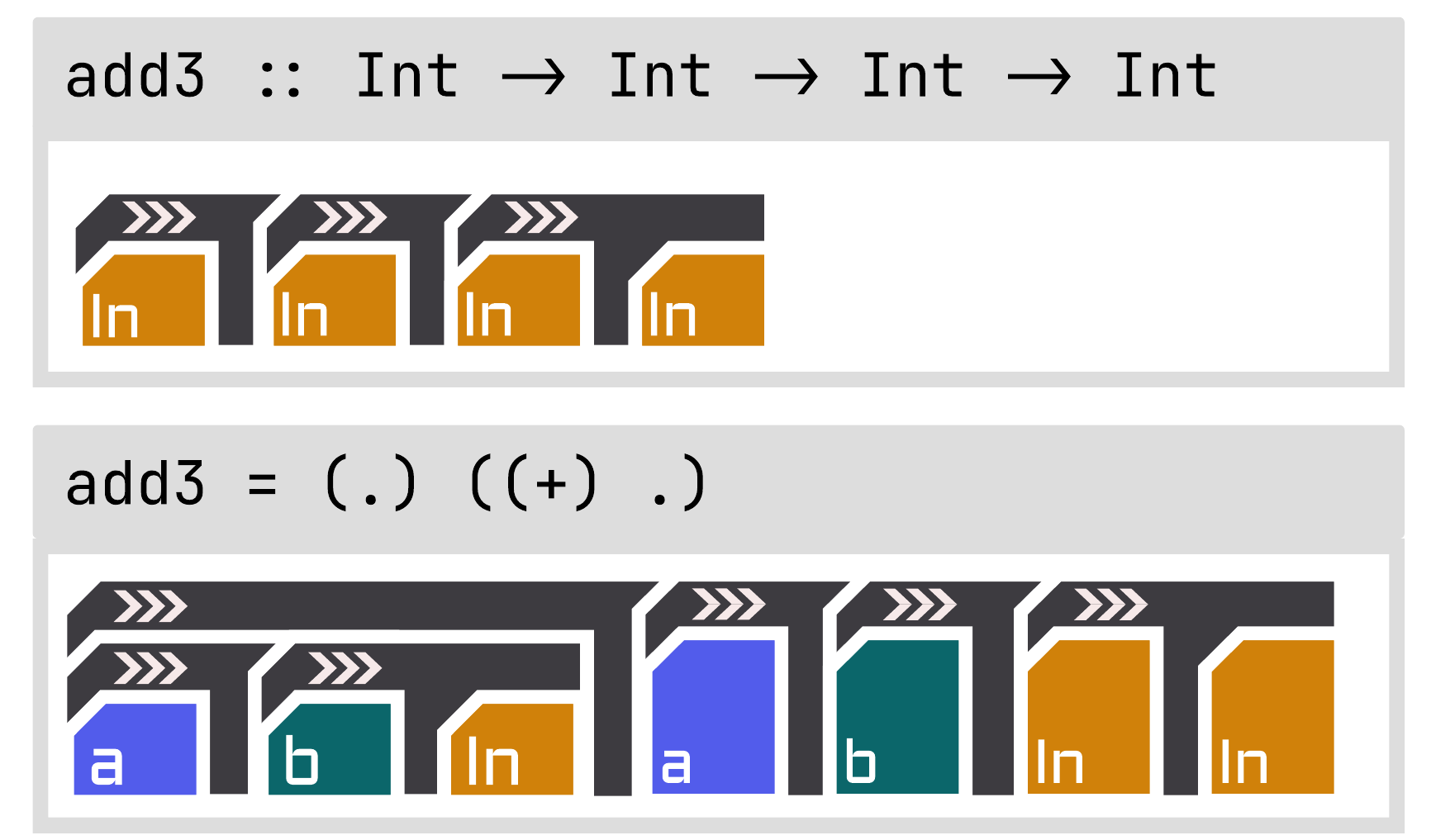}
  \caption{\label{fig:add3} An implementation of function \texttt{add3} but the author missed an (+) from the correct implementation (.) ((+) .) (+). GeckGraph highlights the difference in arity, and reveals that a binary function is needed on the right-hand side for the arity to match. }
\end{figure}

\textbf{The horizontal axis of GeckoGraph} often becomes intuitive when identifying differences in function arities. For example, in Fig. \ref{fig:add3}, the programmer intended to implement a function that sums 3 integers. In the implementation, the programmer missed a \texttt{(+)} function at the end; the resulting function type is largely different in length. It is also clear that the function needs to apply to one more binary function to satisfy the length requirement.  

\begin{figure}[hbt]
  \includegraphics[width=0.6\linewidth]{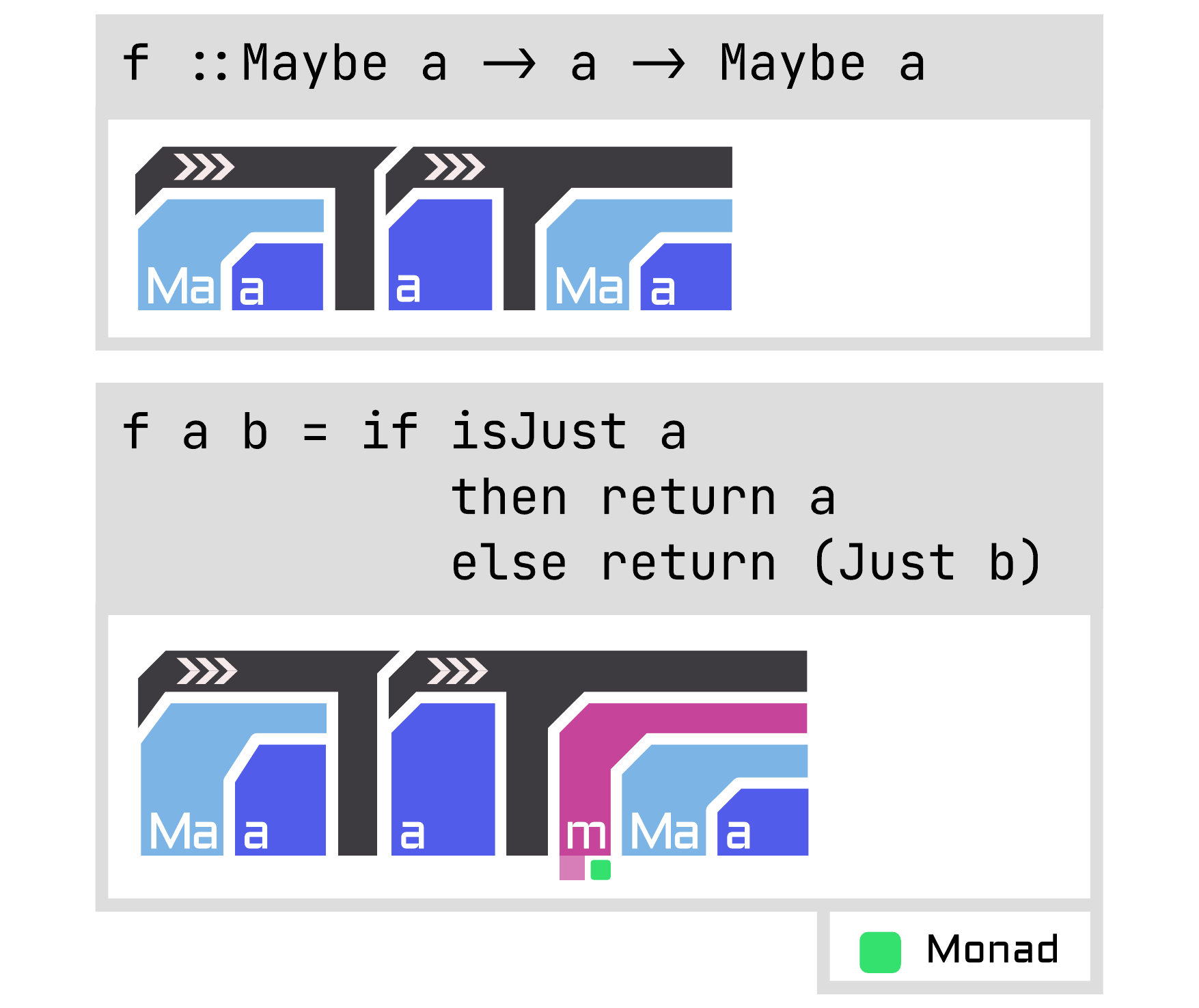}
  \caption{\label{fig:maybe} The function \texttt{f} is planned to have the type \texttt{Maybe a -> a -> Maybe a}. The programmer mistakenly applied the result to the \texttt{return} function, making the result inside a Monad instance.  GeckoGraph reveals the difference in the ``layers" of types. }
\end{figure}

\textbf{The vertical axis of GeckoGraph} often sheds light on the most complex structure of this type. This can often be very useful when inspecting mismatching type errors where data types are nested. Common examples include when programmers forget to apply the value to ``return" in a monadic block or to use \texttt{liftIO} to cast an \texttt{IO} effect. For example, in Fig. \ref{fig:maybe}, the uses of \texttt{return} are excessive. It can be easily identified by examining the difference in the vertical layers of the two types. In text-based type notation, this is distinguished by different pairs of parenthesis. However, parenthesis is an overloaded syntax in type notation. In Haskell, parentheses are used to enclose tuples \texttt{(a, b)}, specify the fixity \texttt{ (a -> b) -> c}, or have no effect \texttt{a -> (b -> c)}.

To quote some feedback from participants: ``Types are much easier to understand by the GeckoGraph than by trying to parse all parentheses and understand the types from the signature. " ``It makes it easier to see at a glance when your output type is correct or what the difference between the current type and the target is."
	
\subsubsection{Low barrier to learn and understand}
GeckoGraph has some key similarities to traditional text-based type notation. GeckoGraph respects the left-to-right reading order. GeckoGraph uses the familiar symbolic name as the secondary encoding. GeckoGraph simulates the prefix notation in type constructors and the infix notation in type operators. With these considerations, we ensured that programmers were able to use GeckoGraph fluently with a minimal amount of training. 

One important class of feedback from the open section is that many programmers mentioned that they did not have any prior knowledge of Haskell but were able to solve the puzzles with the help of GeckoGraph.
``It is similar enough to traditional types that it is intuitive." ``This was how I parse the textual representation of types" was pointed out by multiple participants.

\subsection{Weaknesses}
\subsubsection{Space Usage}\label{subsec:space}
GeckoGraph uses horizontal space in proportion to the size of the type signature syntax tree, and GeckoGraph uses vertical space in proportion to the depth of the syntax tree. Compared to the traditional text-based language, GeckoGraph has the limitation of requiring vertical space. We have identified some approaches to minimize space usage while retaining most of the advantages of using GeckoGraph, such as displaying only the color blocks without the secondary encoding of identifier names.

\subsubsection{Color Encoding}
GeckoGraph highly relies on color hue as a main encoding. It provides a strong visual grouping  \cite{Zeng2023-jz}  for programmers to identify subtle patterns in types, such as the order and placement of substructures. However, the perception of color is different from person to person. This becomes an even bigger issue for color-blind or visually impaired programmers. Although GeckoGraph uses color-blind friendly schemes, it is only a method to avoid indistinguishable types and is not a strong guarantee of effectiveness. For this, we are exploring different encodings, such as patterns and shapes, to maximize the accessibility of GeckoGraph.

\subsection{Gamified Human Study}

It is important to recognize that the human study is designed to be a series of puzzles. The tasks are meant to contain entertaining values. We practiced multiple gamification techniques: levels, story/theme, and goals/rewards. \cite{Hamari2014-mc} This not only allowed us to have confidence that participants are motivated to complete the tasks, it also lent us popularity in the Haskell community and led to a historically high participation rate. Gamification has been shown to improve engagement and motivation. This has been harnessed by many research projects to improve participation in human studies \cite{He2014-vp}. We identify that studies on functional programming are often technical and intimidating; our use of gamification not only attracted historically high participation, but also attracted a wide distribution of experience levels. 

\subsection{Potential Applications}
\subsubsection{Programming Assistance}
We envision many ways GeckoGraph can be integrated into programming tools. GeckoGraph can visualize and inspect types in tooltips and pop-ups. It can be used to discover the mismatching parts of two conflicting types in type errors. It can be used to generate type expressions and edit type expressions structurally. In our post-study survey, the potential integration of text editors and programming assistance were the most requested use cases proposed by the participants. 

\subsubsection{Documentation Assistance}
From what we have learned from our human study, GeckoGraph is well suited to support the documentation of the programming library and the API documentation. It works in tandem with the traditional text-based language and can be generated mechanically, making it possible to standardize with minimal effort. For documentation sites that allow searching by name (e.g., Hoogle \cite{Mitchell_undated-fb}), programmers often need to sieve through a list of identically named functions. For example, a simple Hoogle search for the name \texttt{make} shows a list of functions with vastly different usage and purpose. GeckoGraph can help speed up the selection process by providing a visual notation for each type, and programmers can use a visual grouping of colors, sizes, and positions to home in on the correct documentation page.

\subsubsection{Pedagogical Applications}
We believe that GeckoGraph can be a valuable tool in teaching techniques and theories in programming languages that are difficult to convey in plain language. In fact, many participants in our study reported that they had no prior knowledge of Haskell programming and that they could understand the programming concepts in the game and complete all the puzzles with the help of GeckoGraph.

\begin{figure}[hbt]
  \includegraphics[width=\linewidth]{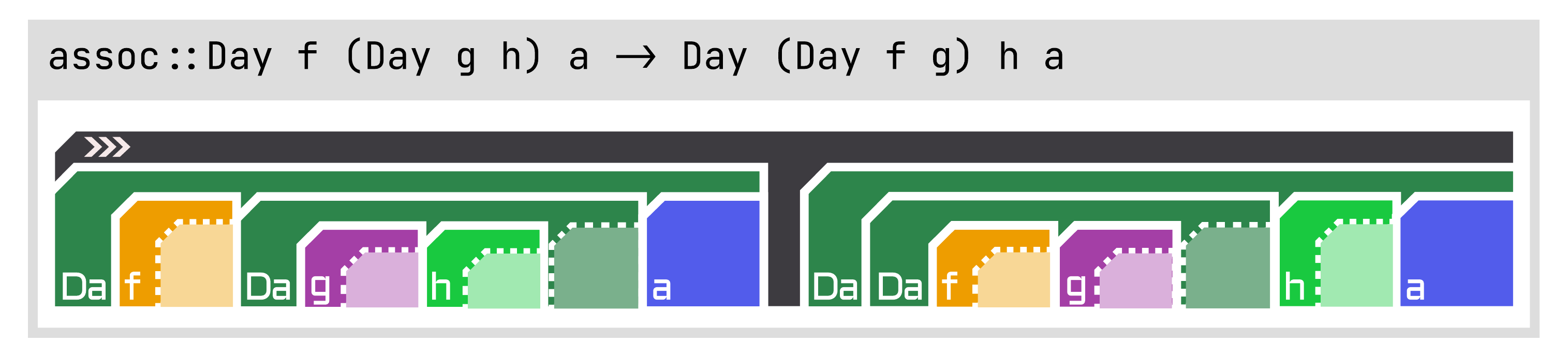}
  \caption{\label{fig:assoc} The \texttt{assoc} function for day convolution \cite{Day1970-kb} in the Kan extension. Even for people who are not familiar with the exact definitions, it is easy to see that the variables \texttt{f},  \texttt{g}, and \texttt{h} are all high-kinded types.}
\end{figure}

Furthermore, the advanced features of GeckoGraph (Section~\ref{sec:benefits}) are also suitable for teaching and learning high-level functional programming concepts. Consider the \texttt{assoc} function for day convolution \cite{Day1970-kb} in the Kan extension (Fig. \ref{fig:assoc}). Although the type signature is short, it is very difficult to trace the semantics mentally due to the number of variables, and their kinds are not obvious from the text-based notation. GeckoGraph makes understanding the type easier by visualizing the ``hidden" higher-kinded types, revealing all the partially applied data types in play.

% GeckoGraph, as depicted in this paper, uses a single color for each type variable and constant. In fact, to create enough visual contrast, the range of the hue spectrum may not provide enough values. In practice, GeckoGraph can encode type names using two or more colors, which will provide enough combinations.  

\section{Related work}
\subsection{Visualizing polymorphic types}
\begin{figure}[hbt]
  \includegraphics[width=\linewidth]{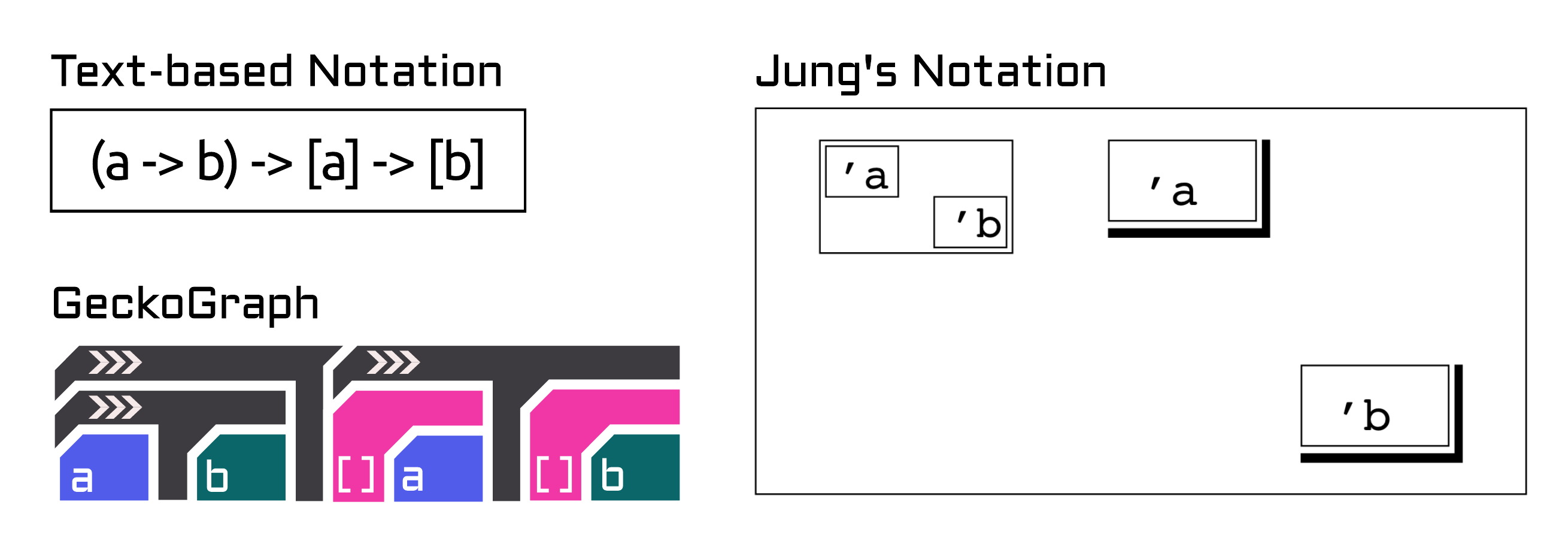}
  \caption{
        \label{fig:jung}
        Comparing the map function (\texttt{(a -> b) -> [a] -> [b]}) in text notation, GeckoGraph, and Jung's notation.
  }
\end{figure}

A similar half-enclosing structure was proposed in the visualization of types by Jung \cite{Jung2000-oc}. In Jung's notation, the type constructor half encloses its arguments, but the figure for the type constructor is placed on the bottom right (Fig. \ref{fig:jung}).  In contrast, GeckoGraph follows the natural reading order, allowing larger structures in a type signature to take precedence over smaller ones. 

Compared to functions in Jung's notation,  GeckoGraph shows two major benefits. First, GeckoGraph naturally translates the general notion of a curried function. Partially, the application of a function can be read as removing the first one of its arguments. This is not the case with Jung's notation. Second, the shape of a function type remains consistent with the shape of normal data types. In Jung's design, a function \texttt{a -> b} looks sufficiently distinct from a data type \texttt{f a b}. This is important because, in functional languages, it is very common for abstraction to be drawn from function and normal data types. For example,  functions and lists both have a functor instance, and their inner values can be altered using a \texttt{fmap} function. The consistent shape of GeckoGraph makes this generalization easier to see visually. 

%Many studies \cite{Jun2000-ec, Jun2000-yu} explored the challenges of how human experts use polymorphic types. The authors then show a history-preserving type inference algorithm and explanation generator that can explain the types in human language. Similarly, \cite{Beaven1993-ay} showed a system that explains the infeasibility of types using human-like languages, explaining the type inference process using a series of "Why" questions and "How" questions. Both studies use natural language to reduce the challenge of understanding polymorphic types. While these studies are promising, explaining polymorphic types with natural language has limitations. They are generally verbose and have to use another visualization assist (text decoration, icons, or indentation) to clarify the naturally hierarchical information. A very similar graphical type representation \cite{Jung2000-oc} that utilizes blocks and colors was proposed. In the paper, the visualization uses 2D blocks to represent types and enclosures to represent type constructors and type arguments. We have discussed thoroughly the difference in design between Jung's notation and GeckoGraph. In addition, the evaluation of Jung's notation uses fixed questions and answers, while GeckoGraph is evaluated using real programming tasks. 

\subsection{Visualization in programming}
Using visualization techniques to improve the comprehension of polymorphic types is not new. This is often practiced to represent document properties, runtime information, and static analysis results.  FluidEdt \cite{Ou2015-vr} displays heap graphs in the left margin. I3 \cite{Beck2015-my} offers search similarity and change history in compact block-based diagrams. Almeida et al. ~\cite{Almeida2022-bv} introduced a novel visualization to aid in understanding ownership and borrowing in the Rust language. While the field of graphical type representation is relatively narrow, it has been studied to some extent. Clerici et al. ~\cite{Clerici2013-ru} proposed a graph-based type inference system that shows the visual representations of unification states. GeckoGraph positions itself similarly to these projects, using color, shapes, symbols, and icons to provide easy-to-glance information. However, GeckoGraph focuses on type-level information, which no other research projects do. In addition, GeckoGraph is evaluated in a much larger study than the other projects, and GeckoGraph is evaluated with a wider range of experience levels. 

\subsection{Visual vs Textual representation}

Many studies have compared the effectiveness of a visual-based programming environment with a textual-based one. Studies \cite{Noone2018-wl, Da_Silva_Ribeiro2014-tm, Cliburn2008-jo, Daly2011-is} found that compared to a purely textual programming language with similar positioning, students who were taught a visual programming language show greater confidence, better retention, and enjoyment in programming courses. While showing a similar trend, GeckoGraph experiments in the context of accompanying text-based notation rather than replacing it.

Many have studied the effect of visual augmentation, providing a visual representation of programming objects without removing the text-based notation. Greenfoot \cite{Montero2010-uh} allows visual and textual representations of programming concepts to be accessible to the learner. PILeT \cite{Alshaigy2015-wy}, providing a programming environment that is an adaptive presentation based on the user's preference. Both tools show positive results in the use of visual augmentation. Although similar to GeckoGraph in combining visual language and text-based programming environment, both studies evaluated the effect based on imperative languages (Java and Python), while our evaluation focused on the effect on a functional language (Haskell).

\section{Conclusion and Future Work}
In this paper, we propose GeckoGraph, a graphical notation for type annotations in functional programming languages. GeckoGraph aims to accompany traditional text-based type notation and to make reading, understanding, and comparing types easier. We conducted a large-scale human study using GeckoGraph compared to text-based type notation, the largest user study on functional programming we are aware of. The results of the study show that GeckoGraph helps improve programmers' ability to succeed in programming tasks we designed, especially for novice programmers.

Our work in this area opens many new directions for future research.  In particular:

\noindent\textbf{In-the-wild Studies}
Although our experiment scenarios are drawn from real-world programming tasks, a certain level of variable control is still applied to remove the effect of familiarity with the tools and libraries. However, it is necessary to assess the usefulness of tools such as GeckoGraph by their real-life usage. To study this, different research methods should be used to study realistic usage and human experience. This may include field deployments or case studies. 

\noindent\textbf{Alternative Type Visualization}
We strongly believe that visualization is an underutilized technique in this effort. GeckoGraph focuses on a faithful view of the tree structure of programming types. However, many more areas and types demand a human-focused approach. For instance, visualizing the ordinal relationship of subsumption or visualizing the numeric changes in dependently typed ``type programs".

\bibliographystyle{ACM-Reference-Format}
\bibliography{paperpile}
\appendix
\section{Game levels} \label{levels}
We provide all the level settings we used in our user study. The online game is still open source and available for evaluation \cite{Anonymous_undated-ne}. However, this can be attempted locally with a Haskell interpreter or even with a pen and paper. The target type is the desired type signature for the function \texttt{zeroToHero}. The available functions show a list of functions that are allowed to be used in the implementation. It is not required to use all the available functions, and it is not forbidden to use any other functions or variables outside the provided functions; even the Haskell prelude is not available.

\subsection{Level 1: Trial run}

\paragraph{Target Type } 
\begin{itemize}
    \item \texttt{zeroToHero :: Zero a -> Hero a}
\end{itemize}

\paragraph{Available Functions} 
\begin{itemize}
    \item \texttt{f :: Zero a -> Hero a}
\end{itemize}

\paragraph{Possible Solution} 
\begin{itemize}
    \item \texttt{zeroToHero z = f z}
\end{itemize}

\subsection{Level 2: Assemble required}

\paragraph{Target Type} 
\begin{itemize}
    \item \texttt{zeroToHero :: Zero a -> Hero a}
\end{itemize}

\paragraph{Available Functions} 
\begin{itemize}
    \item \texttt{runZero :: Zero a -> a}
    \item \texttt{mkHero :: a -> Hero a}
    \item \texttt{(\$) :: (a -> b) -> a -> b}
\end{itemize}

\paragraph{Possible Solution} 
\begin{itemize}
    \item \texttt{zeroToHero z = mkHero (runZero z)}
\end{itemize}

\subsection{Level 3: Which path?}
\paragraph{Target Type } 
\begin{itemize}
    \item \texttt{zeroToHero :: Zero a -> Hero (a, a)}
\end{itemize}

\paragraph{Available Functions} 
\begin{itemize}
    \item \texttt{f1 :: Zero a -> Hero a}
    \item \texttt{f2 :: Zero a -> (a, a)}
    \item \texttt{f3 :: Hero a -> Hero (a, a)}
    \item \texttt{(\$) :: (a -> b) -> a -> b}
    \item \texttt{(.) :: (b -> c) -> (a -> b) -> a -> c}
\end{itemize}

\paragraph{Possible Solution} 
\begin{itemize}
    \item \texttt{zeroToHero z = f3 . f1 \$ z}
\end{itemize}

\subsection{Level 4: A repeating pattern}
\paragraph{Target Type } 
\begin{itemize}
    \item \texttt{zeroToHero :: Zero a b -> Hero b b}
\end{itemize}

\paragraph{Available Functions} 
\begin{itemize}
    \item \texttt{f1 :: Zero a b -> Hero b a}
    \item \texttt{f2 :: Zero a a -> Hero a a}
    \item \texttt{f3 :: Zero a b -> Zero b a}
    \item \texttt{f4 :: Zero a b -> Zero b b}
    \item \texttt{(\$) :: (a -> b) -> a -> b}
    \item \texttt{(.) :: (b -> c) -> (a -> b) -> a -> c}
\end{itemize}

\paragraph{Possible Solution} 
\begin{itemize}
    \item \texttt{zeroToHero z = f2 . f4 \$ z}
\end{itemize}

\subsection{Level 5: A perfect pair}
\paragraph{Target Type } 
\begin{itemize}
    \item \texttt{zeroToHero :: Zero a b -> Hero b b}
\end{itemize}

\paragraph{Available Functions} 
\begin{itemize}
    \item \texttt{fst :: (a, b) -> a}
    \item \texttt{snd :: (a, b) -> b}
    \item \texttt{f1 :: Zero a b -> Hero b a}
    \item \texttt{f2 :: Zero a a -> Hero a a}
    \item \texttt{f3 :: Zero a b -> Zero b a}
    \item \texttt{f4 :: Zero a b -> Zero b b}
    \item \texttt{(\$) :: (a -> b) -> a -> b}
    \item \texttt{(.) :: (b -> c) -> (a -> b) -> a -> c}
\end{itemize}

\paragraph{Possible Solution} 
\begin{itemize}
    \item \texttt{zeroToHero z = snd .f3 . f1 \$ z}
\end{itemize}

\subsection{Level 6: Monty Hall}
\paragraph{Target Type } 
\begin{itemize}
    \item \texttt{zeroToHero :: Zero a b c -> Hero c a}
\end{itemize}

\paragraph{Available Functions} 
\begin{itemize}
    \item \texttt{f1 :: Zero a b c-> Zero c b a}
    \item \texttt{f2 :: Zero a b c -> Zero a c c}
    \item \texttt{f3 :: Zero a b c -> Hero b c}
    \item \texttt{(\$) :: (a -> b) -> a -> b}
    \item \texttt{(.) :: (b -> c) -> (a -> b) -> a -> c}
\end{itemize}

\paragraph{Possible Solution} 
\begin{itemize}
    \item \texttt{zeroToHero z = f3 . f1 . f2 \$ z}
\end{itemize}

\subsection{Level 7: TIE fighter}
\paragraph{Target Type } 
\begin{itemize}
    \item \texttt{zeroToHero :: Zero a b c -> Hero c}
\end{itemize}

\paragraph{Available Functions} 
\begin{itemize}
    \item \texttt{f1 :: Zero a b c -> Hero (a -> b)}
    \item \texttt{f2 :: Zero a b c -> Hero (b -> c)}
    \item \texttt{f3 :: Zero a b c -> Hero a}
    \item \texttt{(<\$>) :: (a -> b) -> Hero a -> Hero b}
    \item \texttt{(<*>) :: Hero (a -> c) -> Hero a -> Hero c}
    \item \texttt{(\$) :: (a -> b) -> a -> b}
    \item \texttt{(.) :: (b -> c) -> (a -> b) -> a -> c}
\end{itemize}

\paragraph{Possible Solution} 
\begin{itemize}
    \item \texttt{zeroToHero z = f2 z <*> (f1 z <*> f3 z)}
\end{itemize}

\subsection{Level 8: The middle man}
\paragraph{Target Type } 
\begin{itemize}
    \item \texttt{zeroToHero :: (a -> d) -> (b -> d) -> (c -> d) -> Zero a b c ->  Hero a d c}
\end{itemize}

\paragraph{Available Functions} 
\begin{itemize}
    \item \texttt{f1 :: Zero a b c -> Zero c a b}
    \item \texttt{f2 :: Zero a b c -> Hero a b c}
    \item \texttt{fmap :: (c -> d) -> Zero a b c -> Zero a b d}
    \item \texttt{(\$) :: (a -> b) -> a -> b}
    \item \texttt{(.) :: (b -> c) -> (a -> b) -> a -> c}
\end{itemize}

\paragraph{Possible Solution} 
\begin{itemize}
    \item \texttt{zeroToHero ad bd cd z = f2  . f1  . f1  . fmap bd  . f1 \$ z}
\end{itemize}

\subsection{Level 9: Split the difference}
\paragraph{Target Type } 
\begin{itemize}
    \item \texttt{zeroToHero :: Zero a b c d ->  Hero d d d d}
\end{itemize}

\paragraph{Available Functions} 
\begin{itemize}
    \item \texttt{f1 :: Zero a b c -> Zero c a b}
    \item \texttt{f2 :: Zero a b c -> Hero a b c}
    \item \texttt{fmap :: (c -> d) -> Zero a b c -> Zero a b d}
    \item \texttt{(\$) :: (a -> b) -> a -> b}
    \item \texttt{(.) :: (b -> c) -> (a -> b) -> a -> c}
\end{itemize}

\paragraph{Possible Solution} 
\begin{itemize}
    \item \texttt{zeroToHero ad bd cd z = f2 \$ f1 \$ f1 \$ f3 \$ z}
\end{itemize}

\subsection{Level 10: The roller coaster}
\paragraph{Target Type } 
\begin{itemize}
    \item \texttt{zeroToHero :: Zero (a -> b -> c -> d) a b c  -> Hero d}
\end{itemize}

\paragraph{Available Functions} 
\begin{itemize}
    \item \texttt{f1 :: Zero (a -> b) a c d -> Zero () b c d}
    \item \texttt{f2 :: Zero a b c d -> Zero b c d a}
    \item \texttt{f3 :: Zero a b c d -> Hero d}
    \item \texttt{(\$) :: (a -> b) -> a -> b}
    \item \texttt{(.) :: (b -> c) -> (a -> b) -> a -> c}
\end{itemize}

\paragraph{Possible Solution} 
\begin{itemize}
    \item \texttt{zeroToHero z = f3 . f2 . f2 . f1 . f2 . f1 . f2 . f1 \$ z}
\end{itemize}

\end{document}